\documentclass[aps,prd,twocolumn,superscriptaddress,longbibliography]{revtex4-2}  

\usepackage{graphicx}  
\usepackage{dcolumn}   
\usepackage{bm}        
\usepackage{amssymb}   
\usepackage[utf8]{inputenc}  
\usepackage[T1]{fontenc} 
\usepackage{amsmath}
\usepackage{color}
\usepackage{hyperref}
\hypersetup{colorlinks=true, citecolor=blue, urlcolor=blue, linkcolor=blue}
\usepackage{chngcntr}

\begin{document}

\title{A roadmap for the design of four-terminal spin valves and the extraction of spin diffusion length}

\author{E. Fourneau}
\affiliation    {Solid-State Physics – Interfaces and Nanostructures, Q-MAT, CESAM, University of Liège,  Liège, 4000, Belgium}

\author{A. V. Silhanek}
\affiliation    {Experimental Physics of Nanostructured Materials, Q-MAT, CESAM, University of Liège, Liège, 4000, Belgium}

\author{N. D. Nguyen}
\affiliation    {Solid-State Physics – Interfaces and Nanostructures, Q-MAT, CESAM, University of Liège,  Liège, 4000, Belgium}

\date{\today}

\begin{abstract}
Graphene is a promising substrate for future spintronics devices owing to its remarkable electronic mobility and low spin-orbit coupling. Hanle precession in spin valve devices is commonly used to evaluate the spin diffusion and spin lifetime properties. In this work, we demonstrate that this method is no longer accurate when the distance between inner and outer electrodes is smaller than six times the spin diffusion length, leading to errors as large as 50\% for the calculations of the spin figures of merit of graphene. We suggest simple but efficient approaches to circumvent this limitation by addressing a revised version of the Hanle fit function. Complementarily, we provide clear guidelines for the design of four-terminal spin valves able to yield flawless estimations of the spin lifetime and the spin diffusion coefficient.
\end{abstract}
\maketitle

\section{Introduction}
The emergence of 2D materials offers new paths for the development of spintronic devices due to their high carrier mobility and low spin-orbit coupling \cite{Avsar2020,Han2014}. In graphene, coherent spin transport of tens of micrometers has been reported \cite{Guimaraes2014,Drogeler2016,Ingla-Aynes2016,Gebeyehu2019}. Currently, efforts are still ongoing to improve the material quality and understand the underlying mechanisms responsible for the spin relaxation phenomenon. Such efforts aim at closing the gap between experimental observations and the theoretical expectation of a 100 $\mu$m spin diffusion length \cite{Huertas-Hernando2009,Han2014,Raes2016,Raes2017}.
\begin{figure}[h!]
\includegraphics[width=0.49\textwidth]{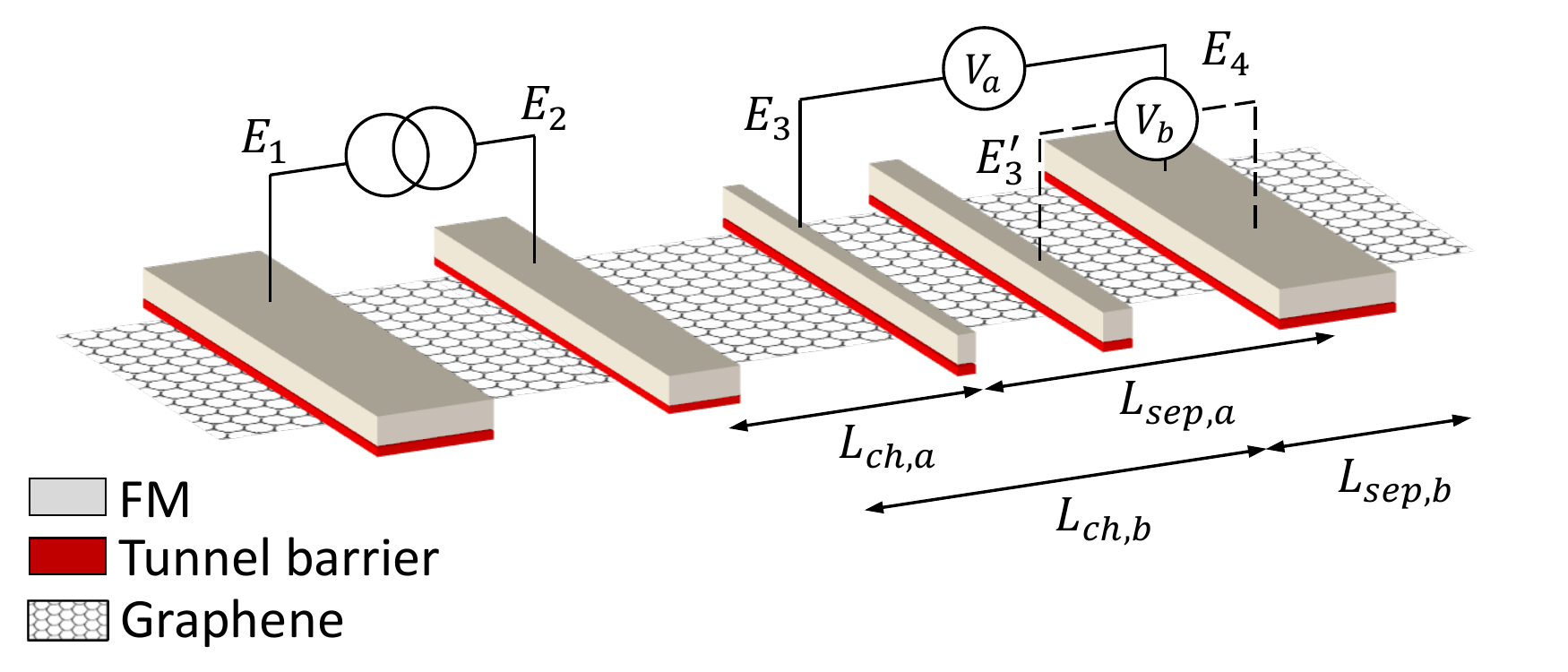}
\caption{Schematic representation of a common spin valve with several FM contacts, allowing spin precession experiment, with different channel lengths (e.g., using electrode E$_3^\prime$ instead of E$_3$) in the same graphene sheet. Two measurement configurations are proposed where the distance between the detector and the outer electrode are different.}
\label{Fig1}
\end{figure}

Hanle spin precession experiment performed in a four-terminals (4T) spin valves geometry (see Fig. \ref{Fig1}) provides an elegant method to evaluate the spin transport properties of a material. In this device, a current is applied between a ferromagnetic (FM) injector (electrode E$_2$) and the reference electrode (E$_1$). Consequently, a spin current is generated under those electrodes which also diffuses towards the detection electrode (E$_3$ or E$_3^\prime$). Therefore, a voltage proportional to the spin accumulation ($\mu_s$) under the detection electrode is probed with respect to the reference electrode (E$_4$). This potential difference $V_{a}=V(E_3)-V(E_4)$ is called the spin voltage. In Hanle experiments, the change of spin voltage as a function of the intensity of a perpendicular magnetic field ($B_\perp$) which induces precession of spin carriers, is measured.  Routinely, experimental results are fitted with the solution of the modified Bloch equations describing diffusive transport in presence of spin precession, naturally leading to two figures of merit (FoMs) of the spin transport: the spin lifetime $\tau_{sf}$ and the spin diffusion coefficient $D_s$ \cite{Johnson1988,Lou2007}. However, this approach presents some limitations as it is based on the following hypotheses:
\begin{enumerate}
\item No spin absorption (contact induced spin relaxation) at any of the contact electrodes,
\item A 1D infinite diffusive medium,
\item An infinite distance between inner (E$_2$, E$_3$) and outer (E$_1$, E$_4$) electrodes,
\item Zero-width electrodes,
\item Non-magnetic (NM) outer electrodes,
\end{enumerate}
The contact-induced spin relaxation has been widely investigated \cite{Volmer2013,Sosenko2014,Volmer2015,Kamalakar2016,Stecklein2016,Amamou2016} and models including the presence of a low resistance injector/detector contacts, as well as the presence of a bias applied to the detector, have been developed \cite{Maassen2012,Idzuchi2015,Gurram2017,Fourneau2020}.
However, limited theoretical efforts have been devoted to understand the effect of geometry despite the large variety of designs and layouts implemented in experimental investigations.
With a finite diffusive medium, specular reflection is expected at transport material edges, leading to an increase of the spin signal \cite{Wojtaszek2014} whereas a short separation between FM inner electrodes and NM outer electrodes induces premature relaxation of spin, leading to a decrease of the global spin lifetime \cite{Drogeler2017,Vila2020}. In addition, the electrode width can be an important feature when its dimension exceeds several hundreds of nm. Spiesser \textit{et al.} \cite{Spiesser2017,Anote} showed that a more accurate theoretical prediction can be obtained by integrating over the injector width. The non-fulfilment of the fifth hypothesis is often observed in spin-valves experiments. In this case, it was shown that the spin signal intensity depends on the magnetic orientation of both inner and outer electrodes \cite{Tombros2007,Jozsa2008,Han2010,Han2012,Neumann2013,Berger2015}. However, quite surprinsingly, there is no general study of its impact on the extracted spin transport FoMs in Hanle precession experiments. Only the case of graphene nano-islands smaller than 1 $\mu$m was analysed \cite{Guimaraes2014-2}. Moreover, no practical solution (neither for experimental nor modeling purpose) has been suggested to correctly account for the modification introduced by the presence of outer FM electrodes. 
Addressing the fourth hypothesis, one can observe that as long as the spin diffusion length ($\lambda_{sf}=\sqrt{\tau_{sf}D_s}$) is short compared with the distance between the contacts ($L_{sep,b}$ in Fig. \ref{Fig1}), the assumption of infinite distances is acceptable since the injected spin signal vanishes before reaching the outer reference electrodes. For higher value of $\lambda_{sf}$, a conflict emerges between the simplification/optimization of fabrication processes and the accuracy on the calculated spin FoMs. Indeed, the most popular method reported in the literature consists in a single lithography step to define multiple contacts followed by the deposition of two successive layers, a tunnel barrier (oxide, h-BN, etc.) and a FM metal. As a consequence, all electrodes can act as a spin injector or detector, with the major benefit of having different channel lengths in a single device depending on the chosen connections. As shown in Fig. \ref{Fig1} one can opt for two configurations: a configuration '$a$' with electrode E$_3$ and a channel length $L_{ch,a}$ and a configuration '$b$' using the electrode E$_3^\prime$ and with a larger channel length $L_{ch,b}$. However, the distance between inner and outer electrodes varies as well with the selected set of electrodes ($L_{sep,a}$ and $L_{sep,b}$ respectively), leading to different extracted FoMs as discussed in this work. 
\begin{figure}[h!]
\includegraphics[width=0.49\textwidth]{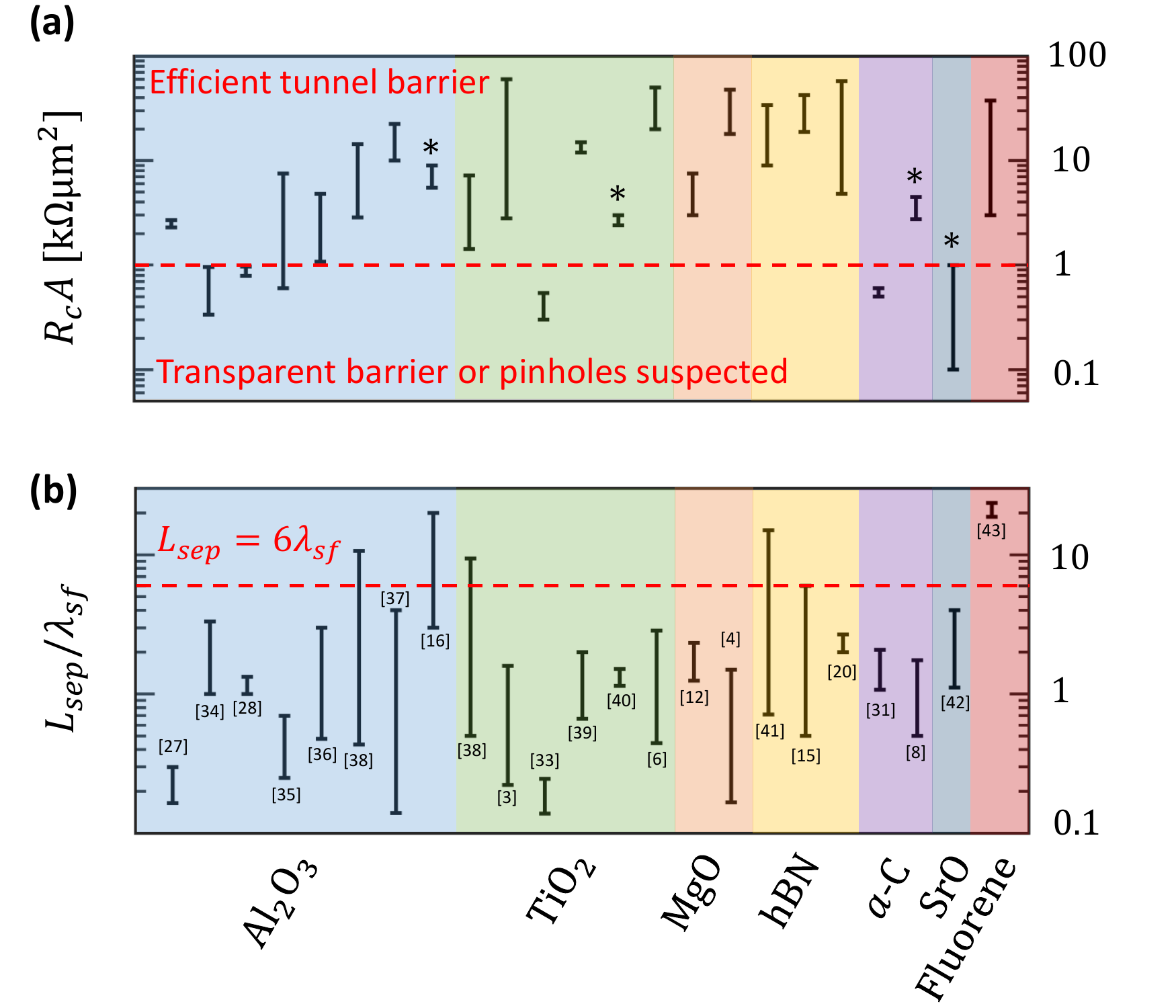}
\caption{Summary of a representative set of graphene-based spin-valves experiments reported in the literature (each bar corresponds to a single scientific publication), for which accurate information relative to the contact resistance, dimension and position is provided. The data range is delimited by the highest and lowest values reported in the corresponding paper. For each barrier material, results are chronologically sorted. Al$_2$O$_3$ \cite{Tombros2007,Tombros2008,Jozsa2008,Popinciuc2009,Jozsa2009,Guimaraes2012,Zomer2012,Stecklein2016}; TiO$_2$ \cite{Zomer2012,Guimaraes2014,Guimaraes2014-2,Kamalakar2015,Torres2017,Gebeyehu2019}; MgO \cite{Volmer2013,Drogeler2016}; hBN \cite{Kamalakar2014,Kamalakar2016,Gurram2017}; $a$-C \cite{Neumann2013,Raes2016}; SrO \cite{Singh2017}; Fluorene \cite{Friedman2014}. (a) Probed contact resistance $R_cA$ for different tunnel barrier materials. The red dashed line corresponds to the lower limit for high quality contact suggested in \cite{Volmer2015}. (b) Ratio of the measured spin diffusion length versus the distance between the inner and the outer detection electrodes. The red dashed line shows the limit of $L_{sep}=6\lambda_sf$ above which negligible effect of outer electrodes is expected.}
\label{Fig2}
\end{figure}

In the light of results reported previously in the literature, it is clear that a great effort has been directed to obtain high quality tunnel barriers, free of pineholes as sumarized in Fig. \ref{Fig2}(a). The vast majority of recent experiments report contact resistance-areas ($R_cA$) larger than the limit of $1$ k$\Omega\mu$m$^2$ suggested by  Volmer et al. \cite{Volmer2015}, whatever the material used as tunnel barrier. On the other hand, all devices reported in Fig. \ref{Fig2} make use of FM outer electrodes except for those with an asterisc \cite{Singh2017,Raes2016,Torres2017,Stecklein2016}. This feature is important because, as shown in panel (b), nearly all devices present a spin diffusion length of the order of the distance between inner and outer electrodes, making the comparison between FoMs of a same set of devices (and therefore between results reported in different papers) inaccurate.
In this work, the impact of the position and material of the outer electrodes is clarified via spin transport simulations in graphene. More importantly, we deduce a series of criteria that must be fulfilled by nominal spin valve devices. Moreover, an updated version of the solution of the modified classical fit function (derived from the solution of the 1D Bloch equations) is proposed in order to include the effect of the distance between electrodes. Even this work focuses exclusively on graphene as diffusive medium, our findings are applicable to any kind of pseudo-substrate with sufficiently high quality spin transport properties. 
\begin{figure*}
\includegraphics[width=0.95\textwidth]{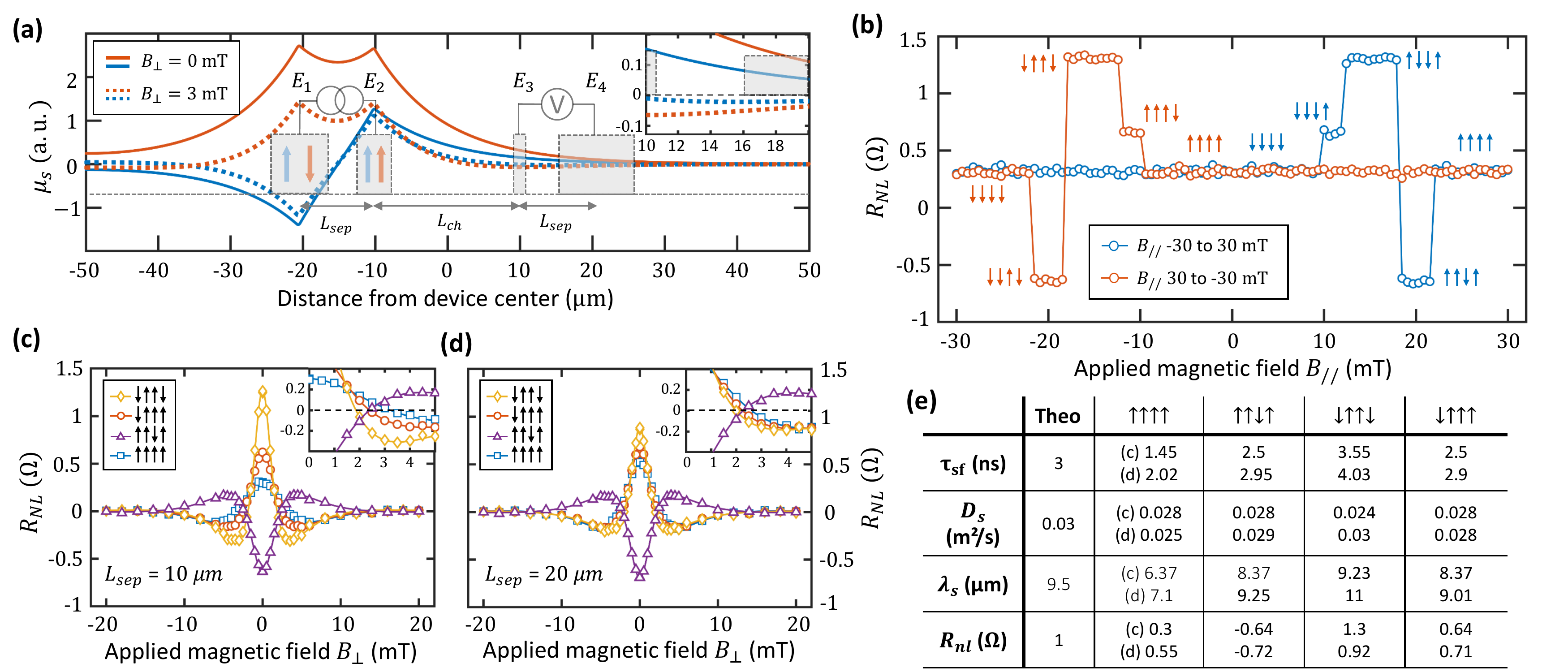}
\caption{Effect of the outer electrodes magnetic orientation and position on the spin FoMs obtained by Hanle precession experiment. (a) Schematic plot of the spin accumulation $\mu_s$ with the position in the device for two opposite directions of the injector magnetization (blue and orange curves) and two external $B_\perp$ values (solid and dashed lines). $\uparrow$ and $\downarrow$ refer to in-plane orientation. Inset is a zoom of the region between E$_3$ et E$_4$. (b) Non-local resistance as a function of the in-plane magnetic field in a range covering all possible magnetic moment configurations. (c-d) Hanle spin precession simulations for a separation distance $L_{sep}$ of (c) 10 $\mu$m and (d) 20 $\mu$m. Solid lines are fitting curves while symbols are simulated data. Insets are zooms on the value $B_\perp$ for which the spin signal is suppressed. (e) Extracted FoMs from Hanle curves (c) and (d). Simulations of (a-d) have been achieved using spin FoMs reported in \cite{Gebeyehu2019} and parameters detailed in Appendix \ref{Appendix:Comsol} as input parameters.}
\label{Fig3}
\end{figure*}

\section{Spin drift-diffusion and Hanle calculations }
In order to demonstrate the importance of the relative magnetic orientation of the outer contact electrodes, calculations have been performed based on data reported in \cite{Gebeyehu2019}. All information regarding the calculation details are developed in Appendix \ref{Appendix:Comsol}. Solving the spin drift-diffusion equations allows one to obtain the spin accumulation profile along the device, as shown in Fig \ref{Fig3}(a) for two relative magnetic orientations for the injector (orange and blue curves). We consider first the case where no $B_\perp$ field is applied (solid line). It is worth noting that the model reproduces the diffusive behaviour of the spin injection theory, with an exponential decay depending on the channel properties, leading to no signal far on the right side of the injector (E$_2$) \cite{Zutic2004}. On the other hand, the signal never reaches zero on the left side as a specular reflection of the spin is assumed on the graphene edges. Since the reference electrode closest to the injector is also FM, it acts as a second spin source ($\downarrow\uparrow$ configuration - blue line) or a spin well ($\uparrow\uparrow$ configuration - orange line). Indeed, as the charge current direction is opposed for both injection contacts, the anti-parallel configuration leads to an amplified signal. Regarding the detection, as the tunnel contacts are resistive enough to avoid spin absorption, the spin accumulation $\mu_s$ is independent of the magnetic orientations and the positions of electrodes E$_3$ and E$_4$.
However, the spin-to-charge conversion factor will be positive for the detecting electrode E$_3$ parallel to the injecting one and negative if anti-parallel. It is shown in Fig. \ref{Fig3}(b), which shows the result of a spin-valves measurement. The in-plane magnetic field $B_{\/\/}$ is swept in order to change the relative orientation of the electrodes as their difference in width leads to a different coercive field. The non-local resistance $R_{NL}$ can acquire four different values depending of the relative orientation of the magnetic moments of the electrodes. Basically, the experimental observation of more than 2 values for $R_{NL}$ is a sufficient proof that external electrodes are perturbing the system while there is still some uncertainty with the opposite observation \cite{Cnote}. 

We focus now on the effect of a spin Hanle experiment ($B_\perp \neq 0$). The FoMS cannot be extracted by fitting Hanle spin precession data with the solution of the modified Bloch equation. Indeed, as shown in Fig. \ref{Fig3}(c), the shape of the Hanle curve depends on the magnetization orientation. In the inset, we observe that the spin signal vanishes for different values of $B_\perp$. This effect is less pronounced if the distance between inner and outer electrodes increases as shown in panel (d) but still persistent. To understand the origin of the change in the spin Hanle curve, calculations of the spin accumulation profile in presence of a perpendicular magnetic field $B_\perp=3$ mT (for which the Hanle signal is close to zero in the configuration $\uparrow\uparrow\uparrow\uparrow$) is displayed in Fig. \ref{Fig3}(a) (dashed lines). Due to spin precession, the spin accumulation reduces more rapidly while diffusing. It results notably in a zero-accumulation point at a specific distance from the injector (see inset for a zoom of the region between the detector and the reference electrode). This position is fully determined by the spin transport properties of the channel and the external magnetic field, and should lead to a root in the Hanle function when the specific position equals the channel length. However, as shown in Fig. \ref{Fig3}(c), this root varies from one configuration of FM electrodes to the other, as the zero-accumulation situation is obtained when $\mu_s(E_3)\pm\mu_s(E_4)=0$, with $'\pm'$ standing for parallel or anti-parallel configurations between E$_3$ and E$_4$, respectively. This deviation of the experimental data results in an inaccuracy for the estimation of $\tau_{sf}$ and $D_s$ when the classical Hanle model is used as fit function. It is worth noting that extracted data look well fitted by the Hanle function as shown in Fig. \ref{Fig3}(c-d). 
\begin{figure}[h!]
\includegraphics[width=0.49\textwidth]{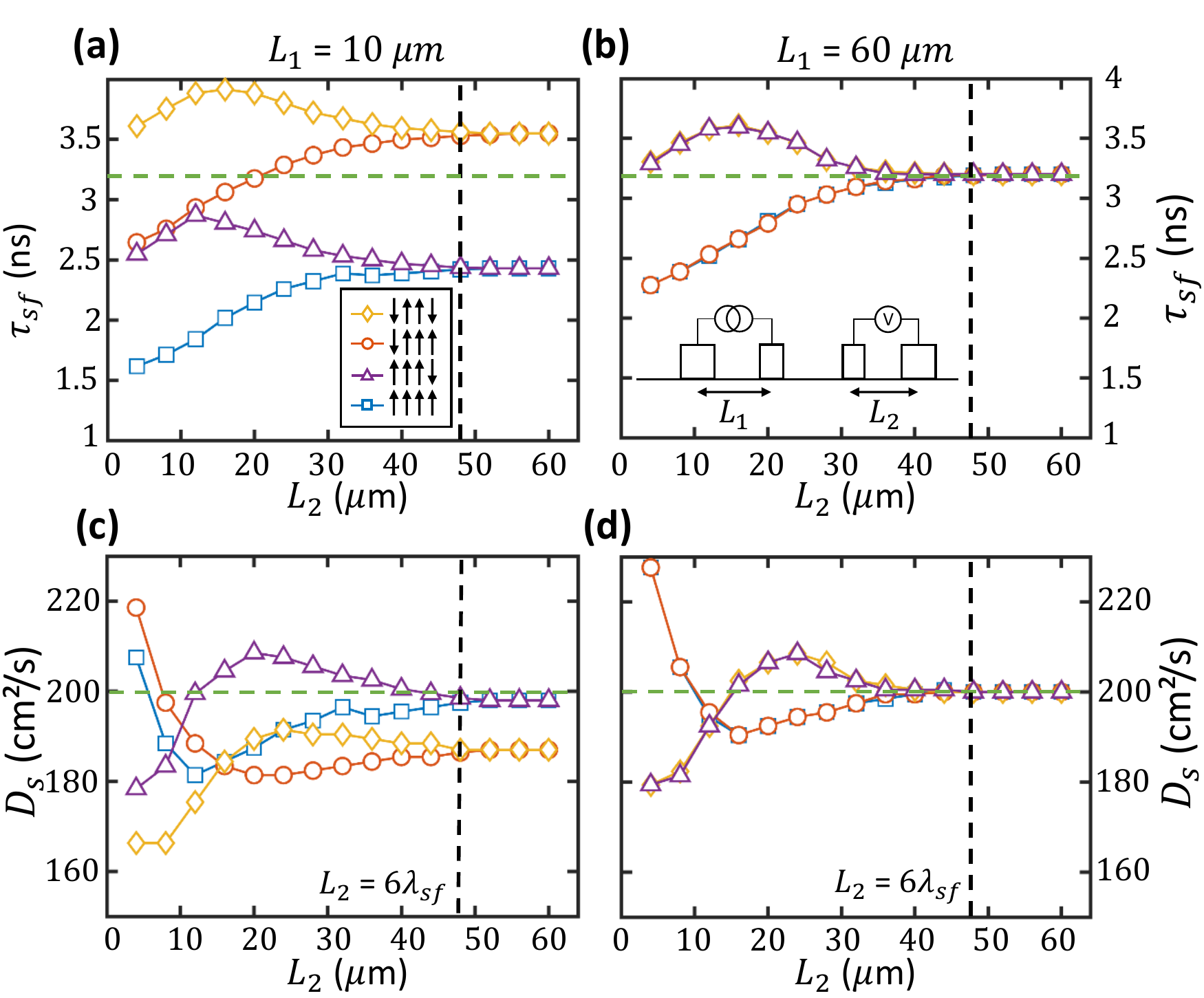}
\caption{Effect of the separation distance $L_2$ between the detector and the closest outer electrode for four different relative magnetizations of the electrodes. Panels (a) and (b) show the extracted spin lifetime $\tau_{sf}$ as a function of $L_2$ for $L_1=10$ $\mu$m and $L_1=60$ $\mu$m, respectively. Panels (c) and (d) are dedicated to the variation of the spin diffusion coefficient $D_s$ in similar conditions. The horizontal green dashed line refers to the theoretical value of $D_s$ and $\tau_{sf}$. The vertical black dashed lines serve as markers for $L_{2}=6\lambda_{sf}$.}
\label{Fig4}
\end{figure}
However, as shown in Fig. \ref{Fig3}(e), measured spin FoMs from panel (c) and (d) depart from the theoretical values used as input for calculations. The diffusion length is underestimated whatever the magnetic configuration, except for the ($\downarrow\uparrow\uparrow\downarrow$) configuration with $L_{sep}=20$ $\mu$m. The spin lifetime is overestimated for both values of $L_{sep}$ in this configuration for which the probed spin signal is maximal. On the other hand, the detected signal is strongly reduced (less than a third of the theoretical value) when the most common configuration ($\uparrow\uparrow\uparrow\uparrow$) is used as both spin sources act oppositely and because the detected voltage is the potential difference between E$_3$ and E$_4$ in the parallel configuration. Regarding the diffusion coefficient, the value is also underestimated but, as it will be shown in Fig. \ref{Fig4}, all these results depend strongly and non-monotonically on the separation distance. It is also shown that configurations ($\downarrow\uparrow\uparrow\uparrow$) and ($\uparrow\uparrow\uparrow\downarrow$) lead to equivalent spin FoMs even through the spin accumulation profile is different (the small difference is due to artificial measurement noise added to the curve). Indeed, as the key parameters are the four distances between injection and detection electrodes (E$_1$-E$_3$, E$_1$-E$_4$, E$_2$-E$_3$, E$_2$-E$_4$), those two specific configurations probe the same signal as long as the separation distance is the same for the injection and the detection part of the device (i.e., E$_1$-E$_3$ is identical to E$_2$-E$_4$). The above observations highlight the importance to include the effect of the outer electrodes when the fifth hypothesis is not fulfilled and therefore when the distance E$_2$-E$_3$ is no longer the only relevant quantity.

In Fig. \ref{Fig4}, the influence of the separation distance at the detector ($L_{2}$) is studied for a fixed separation distance at the injector $L_1$ (in Fig. \ref{Fig3}, $L_1=L_2=L_{sep}$). The channel length is set to 20 $\mu$m. For symmetry reasons, a variation of $L_1$ with fixed $L_2$ leads to the same conclusion. Calculations have been achieved using a spin lifetime of 3.2 ns and a spin diffusion coefficient of 200 cm$^2$/s which are values typically of the order of those reported for graphene based spin experiements \cite{Gebeyehu2019}. Results displayed in panels (a) and (c) are obtained using $L_1=10$ $\mu$m while in (b) and (d) $L_1=60$ $\mu$m. The careful inspection of Fig. \ref{Fig4} leads to several observations. Firstly, results show that both the spin lifetime $\tau_{sf}$ and the spin diffusion coefficient $D_s$ are incorrectly evaluated, and the variation of the inaccuracy with the separation distance is not monotonic for each configuration. Secondly, the magnitude of the associated error depends strongly on the polarization of the FM electrodes. This confirms the observation made previously that the most widely used configuration $(\uparrow\uparrow\uparrow\uparrow$) leads to the largest error for the spin diffusion length. Thirdly, values do converge towards a common reference when the separation distance is larger than nearly $6\lambda_{sf}$, which is therefore considered as the minimal separation distance to ensure a correct fit. As shown in Fig. \ref{Fig2}(b), this criteria is rarely met in the literature. It is worth noting that there are two asymptotic values for data in panels (a) and (c), one for each relative polarization of FM electrode with a separation distance fixed to $10$ $\mu$m. Finally, the horizontal dashed lines represent the theoretical values.  When both separation distances are large enough,  $\tau_{sf}$ and $D_s$ are correctly extracted using the basic form of the fit function.

\section{Method for accurate extraction of spin Figures of Merit}
In order to correctly extract the spin diffusion length, we propose three methods, each of them with pros and cons. A first way is purely based on the fabrication process. The spin-valves device is designed in such a way that the outer electrodes do not influence the Hanle plot. It can be achieved by ensuring that the distance between the inner and outer FM electrodes is large enough ($L_{sep}>6\lambda_{sf}$ leads to less than $1\%$ inaccuracy) or by using non-magnetic tunnel outer electrodes. The obvious downside of these approaches is the requirement of larger devices for the former or more involved nanofabrication processing for the latter (such as a shadow evaporation process \cite{Raes2016} or a double lithography process \cite{Stecklein2016}). An alternative method consists in magnetizing all electrodes at saturation along the same direction and fit the data with a modified version of equation (24) in \cite{Johnson1988}. The new fit function sums the contribution of the four injector-detector couples in a full FM spin-valve:
\begin{widetext}
\begin{equation}
R_{nl}=\dfrac{1}{2}\dfrac{1}{e^2N(E_f)W}\sqrt{\dfrac{\tau_{sf}}{D_s}}\left[F\left(L_{ch},B_\perp\right) - F\left(L_{1}+L_{ch},B_\perp\right)- F\left(L_{2}+L_{ch},B_\perp\right)+ F\left(L_{1}+L_{2}+L_{ch},B_\perp\right)\right]
\label{eq:1}
\end{equation}
\end{widetext}
with $e$ the electron charge, $N(E_f)$ the density of states at the Fermi-level \cite{Bnote} and $W$ the channel width.  The Function $F(L,B)$ is developed in Appendix \ref{Appendix:Fbl}. The major advantage of this approach is that there is no fabrication constraint on the device and therefore, more than four electrodes can be fabricated on one graphene ribbon/flake (for multiple spin-valves with varying channel length as shown in Fig. \ref{Fig1}). It also allows one to work with short separation distance and therefore to drastically reduce the thermal noise. However, as shown in the table of Fig. \ref{Fig3}(e), the non-local resistance is weaker when the configuration $\uparrow\uparrow\uparrow\uparrow$ is used.
The third method consists in performing the Hanle experiment in all four different configurations with a positive zero-magnetic field signal ($\uparrow\uparrow\uparrow\uparrow$, $\downarrow\uparrow\uparrow\uparrow$,$\downarrow\uparrow\uparrow\downarrow$, $\uparrow\uparrow\uparrow\downarrow$) and apply the classical fit method on the mean of those measurements. However, this approach needs specific widths or shapes to obtain a different coercive fields for each FM electrodes. It is worth noting that this method is somehow limited by the fact that it is challenging to ensure an optimal spin injection/detection at each electrodes when they are not magnetized at saturation. 
\section{Conclusion}
In summary, we demonstrate that the distance between inner and outer electrodes of a spin valve device influences strongly the value of extracted spin FoMs when using a Hanle precession method. Our calculations reveal that a separation distance $L_{sep}>6\lambda_{sf}$ is mandatory to avoid the influence of outer electrodes. As this criterion has been hardly met in previous experimental reports, we anticipate that some conclusions concerning the benefits of particular fabrication processes or material choices might need to be revisited in the light of this work. Finally, we demonstrate that all-ferromagnetic spin-valves devices can still provide accurate FoMs if a modified version of the Hanle fit equation (Eq. \ref{eq:1}) is used. 
\section{Acknowledge}
Financial support by the ARC grant 13/18-08 for Concerted Research Actions, funded by the Wallonia-Brussels Federation, is gratefully acknowledged. E. F. gratefully acknowledges the fruitful discussions, shared with W. Keijers, B. Raes and J. Van de Vondel, which served as an inspiration for this work.
\appendix
\renewcommand{\thefigure}{A\arabic{figure}}
\setcounter{figure}{0}

\section{Computational details}
\label{Appendix:Comsol}
Simulations have been performed using the finite element method (FEM) commercial software COMSOL with a geometry based on the cross section of the device presented in Fig. \ref{Fig1}. The charge current distribution has been calculated via the electrical current (ec) module while spin drift-diffusion equations have been manually included for the three spin directions:
\begin{equation}
\begin{split}
&eD\begin{pmatrix}
\nabla n_{s,x}\\
\nabla n_{s,y}\\
\nabla n_{s,z}\\
\end{pmatrix} +
e \mu_n\begin{pmatrix}
n_{s,x}\\
n_{s,y}\\
n_{s,z}\\
\end{pmatrix}\nabla V -\dfrac{g\mu_B}{\hbar}\begin{pmatrix}
n_{s,x}\\
n_{s,y}\\
n_{s,z}\\
\end{pmatrix}\times\begin{pmatrix}
B_{x}\\
B_{y}\\
B_{z}\\
\end{pmatrix}\\  &= \dfrac{e}{\tau_{sf}}\begin{pmatrix}
n_{s,x}\\
n_{s,y}\\
n_{s,z}\\
\end{pmatrix}
\end{split}
\end{equation}
Zero-flux boundary condition has been used at the edge of the graphene and  ribbon. The charge current is fixed at electrode E$_2$ while the electrode E$_1$  is grounded and E$_2$ and E$_3$ are floating. Simulations have been performed using the following parameters: the charge current at the injector $I=5$ $\mu$A, the tunnel contact resistivity $\rho_c=10^{-8}$ $\Omega$m$^2$ (no spin absorption is expected), the carrier density $n=3.6\times10^{12}$ cm$^{-2}$, the spin polarization $P=10\%$, the graphene ribbon width $w=5$ $\mu$m and length $L=200$ $\mu$m (impact of spin carriers reflection at the ribbon edges is negligible), and assuming that the charge and spin diffusion coefficients are equal. 
\section{Spin FoMs correction - practical case}
In order to illustrate why it is important to account for the outer electrodes, we apply our observation on data reported in \cite{Gebeyehu2019} for which geometrical details and magnetic orientation of the four electrodes are given. We compare the classical fit with Eq. (\ref{eq:1}). For the spin lifetime, the value changes from 3 ns using the classical fit to 3.9 ns (increase of 30\%). For the spin diffusion coefficient, the value changes from 0.03 to 0.031 m$^2/$s. It is worth noting that the difference between both fit curves is not visible to the naked eye while the variation of values for the spin FoMs is not negligible.
\begin{figure}[h!]
\includegraphics[width=0.49\textwidth]{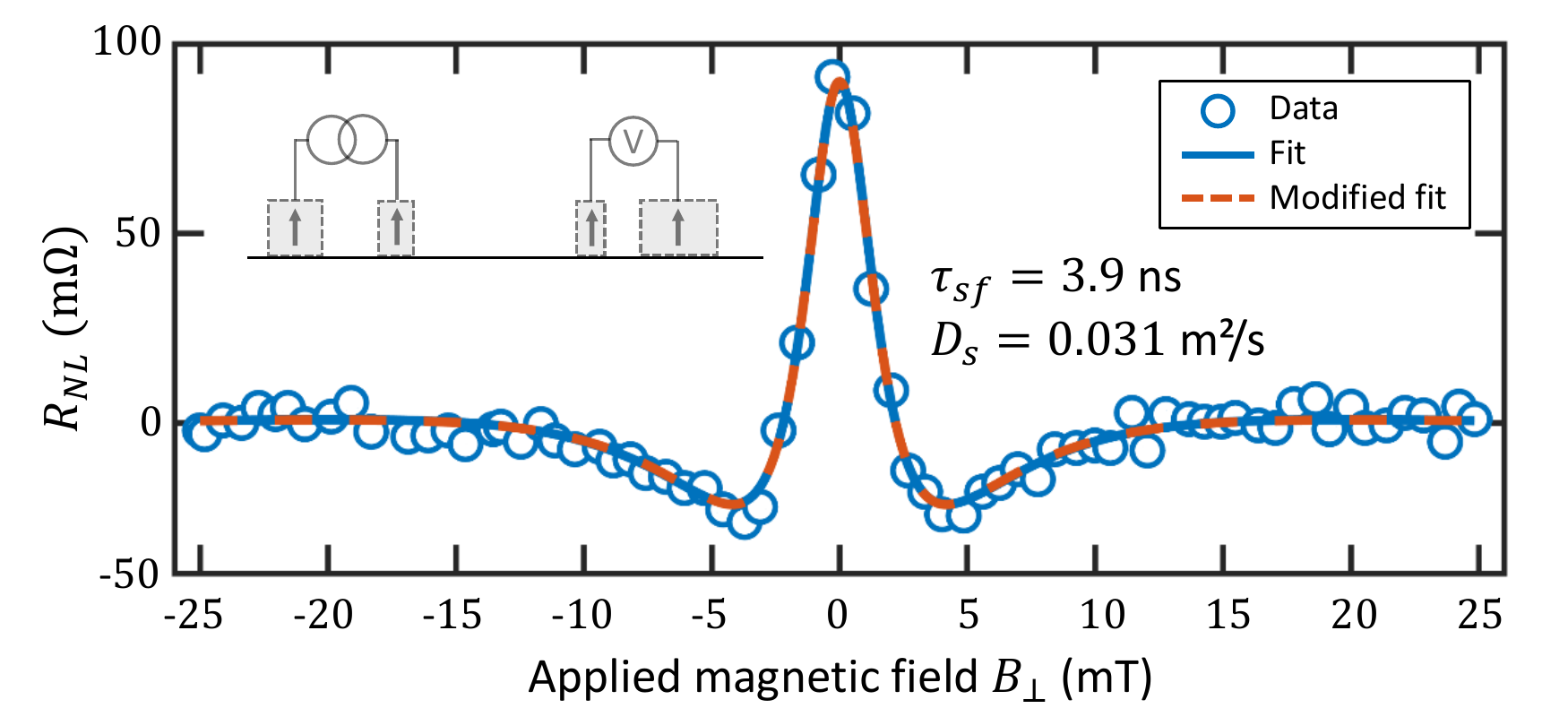}
\caption{Correction of the spin FoMs extraction using Eq. \ref{eq:1} as fit function. Data are from \cite{Gebeyehu2019}.}
\end{figure}
\section{Definition of $F(B,L)$}
\label{Appendix:Fbl}
We define 
\begin{equation}
b=\omega_l\tau_{sf}=-\frac{g\mu_BB_\perp}{\hbar}\tau_{sf},
\end{equation}

\begin{equation}
f(b)=\sqrt{1+\sqrt{1+b^2}},
\end{equation}

\begin{equation}
l=\sqrt{\dfrac{L^2}{\tau_{sf}D_s}},
\end{equation}
where $\omega_l$ is the Larmor frequency, $g$ is the gyroscopic factor, $\mu_B$ is the Bohr's magneton and $\hbar$ is Planck's constant. 
\begin{equation}
\begin{split}
F(b,l)= &\dfrac{1}{f^2(b)-1}\left[f(b)\cos \left(\dfrac{lb}{f(b)}\right)-\dfrac{b}{f(b)}\sin \left(\dfrac{lb}{f(b)}\right)\right]\\
&\times\exp \left(-lf(b)\right).
\end{split}
\end{equation}
\bibliography{biblio}

\begin{thebibliography}{46}
\bibitem [{Cno()}]{Cnote}%
  \BibitemOpen
  \href@noop {} {}\bibinfo {note} {The absence of more than two values for
  $R_{NL}$ in a spin valves measurement excludes the fact that external
  electrodes have no impact only if one can prove that all configurations have
  been probed. For example, only two values for $R_{NL}$ will be observed if
  external electrodes are thin enough to sustain large in-plane magnetic field
  or if each outer electrode has a coercive field similar to the closest inner
  electrode ($H_c(E_1)\simeq H_c(E_2)$ and $H_c(E_3)\simeq
  H_c(E_4)$)}\BibitemShut {NoStop}%
\bibitem [{Bno()}]{Bnote}%
  \BibitemOpen
  \href@noop {} {}\bibinfo {note} {The density of state at the Fermi-Level is
  given by $N(E_F)=2\sqrt{n}/(\hbar v_f\sqrt{\pi})$, assuming a low temperature
  process.}\BibitemShut {Stop}%
\end{thebibliography}
\end{document}